\documentclass[11pt]{amsart}

\title{How to make RSA and some other encryptions 
 probabilistic }

\author{Vitali\u\i\ Roman'kov}
\address{Omsk State  University n.a.  F.M. Dostoevskii}
\curraddr{}
\email{romankov48@mail.ru}

\usepackage{amsmath,amsthm,amsfonts,amssymb}
\usepackage{graphicx}
\usepackage{hyperref}
\usepackage[usenames]{color}

\theoremstyle{definition}

\newcounter{comcount}

\date{}

\begin{document}

\maketitle

\begin{abstract}
A new scheme of probabilistic subgroup-related encryption is introduced.  Some  applications of this scheme based on the RSA, Diffie-Hellman   and ElGamal encryption algorithms are described. Security assumptions and main advantages of this scheme are discussed. We outline that this scheme is potentially semantically secure under reasonable cryptographic assumptions. 
\end{abstract}

\section{Some algorithmic  problems for finite fields and modular rings.} 

In \cite{Rom1}, \cite{Rom2}, we proposed a novel probabilistic public-key encryption, based on the RSA cryptosystem. Its security is based on intractability of the membership and exponent  problems for modular rings and usual security assumptions for the RSA encryption (see exact definitions and explanations below). Also these problems can be considered for finite fields. There are many reasons to estimate these two problems as hard mathematical problems. We note in this connection that the well-known Quadratic residuosity problem  is a partial case of the membership problem.

\subsection{The membership, order and exponent problems for finite fields and modular rings.} 
We consider multiplicative groups  of finite fields and modular rings, specified by a list of generators, or by some other effective way. 
The two most basic questions about such groups are membership in and the order of the group. It is not known how to answer these questions 
without solving hard number theoretic problems (factoring and discrete log). 

Let $n$ be a product of two different primes $p$ and $q.$ Let $\mathbf{Z}_n$ be the corresponding modular ring. Let $Q_n$ be the subgroup of $G_n = \mathbf{Z}_n^{\ast}$ consisting of all quadratic residues. 
The following problem is one of the most known decision problems in number theory and cryptography.

{\bf The Quadratic residuosity problem (QRP).}
 Given an element $f \in G_n$, determine if $f \in Q_n. $
 
 We assume that this determination should   be effective. This problem is considered by many authors as intractable. A number of cryptographic schemes are based on this intractability, 
 and the famous Goldwasser-Micali cryptosystem is one of them. It is important to note that the semantic security property of the Goldwasser-Micali cryptosystem is based on the intractability of the QRP.
The QRP is a particular case of the following decision problem. 
 
{\bf The Membership Problem (MP$_H$)}. Let $G$ be a group, and $H$ be a subgroup of $G.$ Given an element
$f \in G,$ determine if $f \in  H.$

It seems that the following problem was previously considered only for matrix groups.

{\bf The order problem (OP).} Given an element $g$ (subgroup $H$) of a group $G$, determine  order $|g|  \ (|H|). $

{\bf Lemma 1.} Let $p$ and $q$ be two different odd primes such that $p, q \equiv 3(mod \ 4),$ and $n =pq$. Then solvability of OP for 
$G_n=\mathbf{Z}_n^{\ast}$ implies solvability of QRP for $G_n$. 

Proof. Let $p =4k+3$ and $q=4l+3.$ Then  $|G_n|=(p-1)(q-1)=4(2k+1)(2l+1)$. As  index of $Q_n$ in $G_n$ is $4$, then   $Q_n$ has odd order $m=(2k+1)(2l+1).$ An element $g$ lies in $Q_n$ if and only if  $|g|$ is odd. Hence, if we can determine $|g|$ for each $g\in G_n$, we can solve QRP for $G_n$. 

Let $\mathbf{F}_q, q=p^r,$ be the finite field of order $q$ and characteristic $p.$ If the primality decomposition of $q-1$ is known
 then there is a polynomial algorithm solving OP for elements of $\mathbf{F}_q^{\ast}$ (see \cite{Men}). 

 {\bf The exponent problem (EP).} Given a subgroup $H$ of a group $G$, determine exponent $e(H)$, i.e., the minimal positive integer $e$ such that $h^e=1$ for every element $h\in H.$ 
 
 Obviously, the solvability EP implies solvability OP, and by Lemma 1 solvability QRP in the case $p,q\equiv 3(mod \ 4).$

 \subsection{Constructing finite fields and  modular rings that  contain subgroups of prescribed exponent.} 
 Let $r$ be a positive integer, and one party, say Bob, wants to construct a field $\mathbf{F}$ of a sufficiently large size that contains an element $g\in \mathbf{F}^{\ast}$ such that $|g|=r.$ Then he seeks to find a prime $p$ in the form $p=1+2rx$   where $x$ is  taken  randomly. He checks the primality of $p$ with some of the known primality tests. When it turns out $p$ is prime, there is an element $g \in \mathbf{F}_p^{\ast}$ such that $|g| = r$. It can be  found by ordinary effective procedure (see \cite{Men}). As $r$ divides $p^k-1$ for any $k$, he can take  field $\mathbf{F}_q$ for  $q =p^k$ for suitable $k$, and to find element $g\in \mathbf{F}_q^{\ast}$ such that $|g|=r.$ 
 If he wants to get a couple $g_1, ..., g_t$ of elements such that $|g_i|=r_i$, for prescribed positive integers $r_i$ \  ($i=1, ..., t$) he seeks to find a prime $p$ in the form $p=1+2rx$, where $r =\prod_{i=1}^tr_i$, and $x$ is taken randomly. Then he gets desired elements $g_1, ..., g_t$ of a field $\mathbf{F}_q$, $q =p^k$ for a suitable $k$ as above.
 
 Let $e$ be a positive integer, and Bob wants to  construct a field $\mathbf{F}$ of a sufficiently large size such that the multiplicative group  $ \mathbf{F}^{\ast}$ contains subgroup $H$  of exponent $e.$ Then he generates  a couple of positive integers $r_1, ..., r_t$ such that lcm$(r_1, ..., r_k) =e,$ then constructs a field $\mathbf{F}_q, q=p^k,$ for a suitable $k$, and finds elements $g_1, ..., g_t\in \mathbf{F}_q^{\ast}$ such that $|g_i|=r_i$ for $i=1, ..., t,$ as explained above, 
 and then succeeds in determining $H=$ gp$(g_1, ..., g_t).$
 
 Let $r$ be a positive integer, and Bob wants to construct a modular ring $\mathbf{Z}_n$ of a sufficiently  large size $n = pq$, where $p$ and $q$ are two different odd prime numbers,  that contains an  element $g\in \mathbf{Z}_n^{\ast}$  such that $|g|=r.$ Then he seeks to find primes $p$ and $q$ in the forms $p=1+2r_1x$ and $q=1+r_2y$, where $r_1$ and $r_2$ are positive integers such that $r=$ lcm$(r_1, r_2)$, and $x, y$ are taken randomly. He checks the primality of $p$ and $q$ with some known primality test. Then he finds elements $g_1\in \mathbf{F}_p^{\ast}$ and $g_2\in \mathbf{F}_q^{\ast}$ such that $|g_1|=r_1$ and $|g_2|=r_2,$ as above. By the Chinese Remainder Theorem, he gets a solution $g$ of the set of equations $g = g_1(mod \ p), z = g_2 (mod q). $ It follows that $|g|=r.$ Hence Bob can construct a couple $g_1, ..., g_t$ of elements such that $|g_i|=r_i$, for prescribed positive integers $r_i$ \  ($i=1, ..., t$), and to construct a modular ring $\mathbf{Z}_n, n =pq,$ where $p$ and $q$ are different odd primes, that contains a subgroup $H\leq \mathbf{Z}_n^{\ast}$ of prescribed exponent $e(H)=e.$ 
 
 When Bob publics the  field $\mathbf{F}_q, q=p^k,$ or modular ring $\mathbf{Z}_n, n = pq,$ and the subgroup $H=$ gp$(g_1, ..., g_t),$ constructed as above, he doesn't public the primality decompositions of $p-1$ in the field case,  and  keeps secret the factors of $n$ in the modular ring case. In both cases  the exponent $e=e(H)$ is private.  
 
 Oscar, an opponent, who wants to crack, modify, substitute, or replay messages, cannot compute $e$ without solving hard number theoretic primality factoring problem with respect to $p-1$ in the field case. In the modular ring case, he needs not only in knowing of the factors of $n,$ but also  he needs in primality factorings of $p-1$ and $q-1$. Hence, in both the cases,  the exponent problem with respect to $H$ can be considered as intractable. 

 \section{Basic scheme of probabilistic subgroup-related encryption founded on the intractability of EP. } 
 
 Suppose two parties, say Alice and Bob,  want to establish a secure transport connection through a non-secure channel. They agree to use the multiplicative group $K^{\ast}$ as a platform, where $K$ is a finite field
 $\mathbf{F}_q, \ q=p^k,$ or a modular ring $\mathbf{Z}_n, \ n=pq,$ where $p$ and $q$ are different sufficiently large primes. Also they agree that Bob will choose all parameters and  then he will send all public parameters to Alice. The encryption by Alice, and the decryption by Bob will be  done as follows .
 
 \bigskip
  {\bf Field version.}
 
 \begin{enumerate}
 \item Bob creates a probabilistic cryptographic system. Namely, he chooses a field $\mathbf{F}_q, \ q=p^k$ and two subgroups $H$ and $U$ of $\mathbf{F}_q^{\ast}$ of coprime orders $r$ and $s$ respectively. Elements of $U$ encode all possible messages, and elements of $H$ play role of masks. Also Bob computes $t =r^{-1}(mod \ s).$
 \item Bob sends the public parameters, namely: $p, k, H, U$, to Alice. The subgroups $H$ and $U$ are specified by their generating elements, or by some other effective method that does not reveal the secret parameters $r$ and $s$.   
 \item To transport a message $u\in U, $ Alice chooses $h\in H$ randomly, then she sends $g =hu$ to Bob through non-secure channel. 
 \item Bob receives $g.$ Then  he succeeds to reveal the message $u$ as follows:  
 $$g^{rt}=(h^r)^t(u^{rt})= u.$$
 \end{enumerate}

 {\bf Remark 1.} 
 \begin{itemize}
 \item In particular, the subgroups $H$ and $U$ can be chosen such that $\mathbf{F}_q^{\ast}=HU.$ Then every element $g\in \mathbf{F}_q^{\ast}$ can be  uniquely written in the form $g=hu,$ where $h\in H$ and $u\in U.$
 \item  The orders $|H|$ and $|U|$ have to be sufficiently large. The presentations of $H$ and $U$ should be chosen in  a way that  doesn't give a possibility  to find the decomposition  $n=pq$.
 \end{itemize}
 
 \bigskip
 {\bf Modular ring version.}
 
  \begin{enumerate}
 \item Bob creates a probabilistic cryptographic system. Namely, he chooses a modular ring $\mathbf{Z}_{n}, \ n=pq, $ $p$ and $q$ different primes, and two subgroups $H$ and $U$ of $G_n=\mathbf{Z}_n^{\ast}$ of coprime orders $r$ and $s$ respectively. Elements of $U$ encode all possible messages, and elements of $H$ play role of masks. Also Bob computes $t =r^{-1}(mod \ s).$
 \item Bob sends the public parameters, namely: $n,  H, U$, to Alice. The subgroups $H$ and $U$ are specified by their generating elements, or by some other effective method that does not reveal the secret parameters $r$ and $s$.   
 \item To transport a message $u\in U, $ Alice chooses $h\in H$ randomly, then she sends $g =hu$ to Bob through non-secure channel. 
 \item Bob receives $g.$ Then  he succeeds to reveal the message as follows. 
 $$g^{rt}=(h^r)^t(u^{rt})= u.$$
 \end{enumerate}

 {\bf Remark 2.} 
 \begin{itemize}
 \item In particular, the subgroups $H$ and $U$ can be chosen such that $\mathbf{F}_q^{\ast}=HU.$ Then every element $g\in \mathbf{F}_q^{\ast}$ can be  uniquely written in the form $g=hu,$ where $h\in H$ and $u\in U.$
 \item  The orders $|H|$ and $|U|$ have to be sufficiently large. The presentations of $H$ and $U$ should be chosen in  a way that  doesn't give a possibility  to find the decomposition  $n=pq$.
 \end{itemize}

\section{Some applications.}

\subsection{Probabilistic encryption based on  RSA encryption.}

The following encrypting has been proposed in \cite{Rom1} and \cite{Rom2}. The described above basic scheme is combined with  the standard RSA algorithm.

Let $p$ and $q$ be two different odd primes. Bob chooses subgroups $H$ and $U$ of the multiplicative group $G_n = \mathbf{Z}_n^{\ast}$ of the modular ring $\mathbf{Z}_n$ such that  
their respective orders $t$ and $r$ are coprime. He presents these subgroups by their generating elements as: $U=$ gp$(u_1 ,..., u_m)$ and $H=$ gp$(h_1 ,..., h_s),$ or in a different effective way. 
We suppose that $U$ is the message space, i.e., each message $u$ is presented as an element of $U$, and vice
versa. Now we fix public and secret data that are established off-line as follows. 

Public  data: $n$ (and therefore $\mathbf{Z}_n$ and $G_n = \mathbf{Z}_n^{\ast}$), $u_1 ,...,u_m$ (and therefore $U$), $h_1,..., h_s$ (and therefore $H$).

 Secret data: $p, q, \varphi (n) = (p - 1)(q -1 ), r, t.$
 
 Alice chooses public key $e \in \mathbf{Z}$ such that gcd$(e, r) = 1$. Then she computes the secret key $d = td_1$ such that 
  $(te)d_1 = 1 (mod \  r). $ This is possible because gcd$(t, r) =$ gcd$(e, r) = 1$ by our assumption. Then
$ted = 1 + rk $ for some integer $k$. Thus, Bob has the following keys:

Public key: $e.$
 
 Secret key: $d.$
 
 To send a message $u \in U$  to Bob, the other party, Alice, chooses a random element $h \in H$ (secret
session key) and acts as follows:
 
 Encryption: $u \rightarrow c = (hu)^e (mod \ n).$

Bob recovers $u$ as follows: 
 
 Decryption: $c \rightarrow u = c^d (mod n).$
 
 Correctness: $c^d= (h^t)^{ed_1} u(u^r)^k = u (mod \ n).$
 
 Security assumption. If the adversary can to find the factors $p$ and $q$ of  $n$ he can not reveal message $u$ without knowing of primality factorings of $p-1$ and $q-1$. Hence, a security of the algorithm is based on intractability of the full primality decomposition problem for  positive integers. A priori, the primality decomposition problem is harder than the decomposition problem for positive integers of the form $n = pq.$

 \subsection{Probabilistic encryption based on the ElGamal key exchange protocol.} The ElGamal algorithm \cite{EG} is an asymmetric encryption algorithm for public key cryptography, based on hardness of the Discrete logarithm problem. ElGamal is semantically secure under reasonable assumptions, and is probabilistic \cite{GM}.
 The original ElGamal encryption scheme  \cite{EG} operates as follows. First, Bob fixes a large prime $p$, a generating element  $g$ of the multiplicative group $\mathbf{F}_p^{\ast},$ and a positive integer $a$ in the range $1 < a < p$.  Bob's private key is taken to be $a,$ while his public key is the tuple $(g,y=g^a,p).$
Alice performs encryption by segmenting the plaintext  and encoding it as a sequence of integers in the range 
 $0 < u < p.$  Alice chooses a
temporary secret in the form of an auxiliary random integer $r$ and encrypts a plaintext as
$c = u \cdot y^{r}=u\cdot g^{ar}$.
Along with this encrypted message $c$, Alice includes the header $g^r$. 

Bob performs decryption by first manipulating the header $(g^r)^a=g^{ar}.$
It follows that the original message can be recovered by noting that
$c\cdot (g^{ar})^{-1} = u.$

The ElGamal cipher leverages the purported difficulty of
computing the discrete logarithm. 
 
 We propose a new version of the ElGamal encryption that does not use  header. As the original,  our version is probabilistic, but uses a very different idea. This encryption scheme operates as follows. First, Bob  chooses two positive coprime integers $r$ and $s$. Then he  fixes a large prime $p$ such that $p-1$ divides to  $rs$, i.e., $p-1=rsk$ for some $k.$
 He sets $a=sk$ and $b=rk.$ Bob's private key is taken to be the tuple $(r,s,a)$ while his public key is the tuple $(g,y=g^a,g^b,p).$

 He announces that the message space is subgroup gp$(g^b)$ which order is $s.$  Then he computes $t=r^{-1}(mod \ s).$
 
 Alice chooses a
temporary secret in the form of an auxiliary random integer $l$ and encrypts a plaintext $u = (g^b)^x=g^{bx}$ as
$c = u \cdot y^{l}=u\cdot g^{al}$.

Bob performs decryption by first manipulating $c^r=u^r\cdot (g^{al})^r=u^r\cdot (g^{rsk})^l =u^r\cdot (g^{p-1})^l=u^r.$ It follows that the original message can be recovered by noting that $(u^r)^t=u.$ 
 
Let $u_1, u_2 \in $   gp$(g^b)$  be two different messages, and $c_i=u_i\cdot g^{al} \  (i\in\{0,1\}$ be the cipher text of one of them.
The Oscar, opponent, has to guess about correct value of $i.$ Suppose  that he can give correct answer with probability significantly greater than $1/2.$ As $u_j^{-1}\cdot c_i$ lies in gp$(g^a)$ if and only if $i=j,$ Oscar can to solve the membership problem for gp$(g^a)$ with  probability significantly greater than $1/2.$
Assuming the latter problem is hard when we take  elements from  gp$(g^a)\cdot $gp$(g^b)$, we conclude that this version of ElGamal encryption is semantically secure. Note that we can take $a$ and $b$ such that 
gp$(g^a)\cdot $gp$(g^b)= \mathbf{F}_p^{\ast}.$
 
 \subsection{Probabilistic encryption based on the Diffie-Hellman key exchange protocol.} 

The Diffie-Hellman algorithm \cite{DH} is an asymmetric encryption algorithm for public key cryptography, based on hardness of the Discrete logarithm problem.  The original Diffie-Hellman encryption scheme  \cite{DH} operates as follows. First, Alice and Bob agree upon a large prime $p$, number $q=p^t$ and a generating element  $g$ of the multiplicative group $\mathbf{F}_q^{\ast}.$   Bob's private key is taken to be a number $b$ while his public key is  $g^b.$
Alice's private key is a number $a$  and her public key is $g^a$. The secret key they exchange is then $g^{ab}.$ Both users can compute this key. An unauthorized third party will be unable to determine the key if it is computationally infeasible to compute $g^{ab}$ knowing only $g^a$ and $g^b.$

We propose a new version of the Diffie-Hellman encryption that  is probabilistic. This encryption scheme operates as follows. First, Bob chooses  a private positive  integer $r$, and Alice chooses  a private positive integer $s$. Then Bob takes randomly number $x$ and publics  $r_1=rx.$ Alice takes randomly $y$ and publics $s_1=sy$. Then they agree upon prime of the form $p=1+2r_1s_1z.$ They use field $\mathbb{F}_p$ as platform and agree upon some fixed public element $g\in \mathbb{F}_p^{\ast}$. Ideally, $g$ should be a generator of $\mathbb{F}_p^{\ast}$; however, this is not absolutely necessary. Bob finds a subgroup $H$ of $\mathbb{F}_p^{\ast}$ of exponent $r$ and publics its generators;   Alice finds a subgroup $U$ of exponent $s$ and publics its generators.  

To exchange secret key Bob chooses randomly private number $b$, element $u\in U,$ and Alice chooses randomly private number $a$ and element $h \in H.$ Bob sends $ug^{br}$ to Alice, 
Alice sends $hg^{as}$ to Bob. 

Bob computes $(hg^{as})^{rb}=g^{abrs}.$ Alice computes $(ug^{br})^{sa}=g^{abrs},$ that is the exchanged secret key.

\end{document}